\documentclass[journal]{IEEEtran}

\usepackage{color,array,amsthm}
\usepackage{graphicx}
\usepackage{amsmath}
\usepackage{amsthm}
\usepackage{amssymb}
\usepackage{booktabs}
\usepackage{subcaption}

\usepackage{url}

\newcommand{\Lim}[1]{\raisebox{0.5ex}{\scalebox{1}{$\displaystyle \lim_{#1}\;$}}}

\begin{document}

\title{Extended Road-Aware Line-of-Sight Probability Model for Urban Air Mobility} 

\author{Abdullah~Abu~Zaid,
        Baha~Eddine~Youcef~Belmekki,
        and~Mohamed-Slim~Alouini
\thanks{Authors’ addresses: Abdullah Abu Zaid and Mohamed-Slim~Alouini are with the King Abdullah University of Science and Technology, Thuwal 23955, Saudi Arabia (e-mail: \{abdullah.abuzaid, slim.alouini\}@kaust.edu.sa). Baha Eddine Youcef Belmekki is with the School of Engineering and Physical Sciences, Heriot-Watt University, Edinburgh EH14 4AS, United Kingdom. Email: b.belmekki@hw.ac.uk. \textit{(Corresponding author: Baha Eddine Youcef Belmekki)}.}}

\maketitle

\begin{abstract}
As urban air mobility (UAM) emerges as a transformative solution to urban transportation, the demand for robust communication frameworks capable of supporting high-density aerial traffic becomes increasingly critical. An essential area of communications improvement is reliably characterizing and minimizing interference on UAM aircraft from other aircraft and ground vehicles. To achieve this, accurate line-of-sight (LOS) models must be used. In this work, we highlight the limitations of a LOS probability model extensively used in the literature in accurately predicting interference caused by ground vehicles. Then, we introduce a modified probability of LOS model that improves interference prediction by incorporating the urban topography and the dynamic positioning of ground vehicles on streets. Our model's parameters are derived from extensive simulations and validated through real-world urban settings to ensure reliability and applicability.
\end{abstract}


\section{INTRODUCTION}
The rapid pace of technological innovation enabled the advent of advanced air mobility (AAM) as a transformative mode of future transportation, which will reshape how we navigate urban landscapes. Central to AAM is the development of electric vertical takeoff and landing (eVTOL) aircraft, which are designed to operate in urban environments, offering an innovative solution to reduce surface-level congestion and contribute to the reduction of urban air pollution. Urban air mobility (UAM), enabled by eVTOLs and considered a subset of AAM, is envisioned as a promising solution to the challenges of urban transportation, namely congestion fueled by rapid urbanization and the limitations of traditional ground transportation infrastructure \cite{UAM2018NASA}. Practical applications of UAM include autonomous cargo delivery services, which enhances logistical efficiency, and passenger air taxis, which offer high-speed urban travel \cite{pons2022understanding}. As UAM becomes an essential element of modern urban transportation, the need to develop sophisticated communication infrastructure that can support the operation of eVTOL aircraft and their air corridors becomes evident \cite{greenfeld2019concept}. The integration of eVTOL aircraft into urban transportation introduces complex communication challenges that must be addressed to ensure safety, efficiency, and reliability \cite{zaid2023evtol}.

Non-terrestrial networks (NTNs) are emerging as promising solutions for providing aerial coverage to both UAM and ground-level users, offering expansive coverage to UAM aircraft and maintaining nearly constant line-of-sight (LOS) connections with these aerial vehicles \cite{zaid2023aerial, huang2024system}. NTNs can be categorized into three main types: space-based, which include satellites \cite{kodheli2020satellite}; stratospheric-based, such as high-altitude platform stations (HAPS) operating at altitudes of 17-23~km \cite{belmekki2024cellular, lou2023haps}; and tropospheric-based, such as networked tethered flying platforms (NTFPs) flying between 200~m to 4~km. Each category of NTN offers unique advantages and is expected to play a crucial role in providing connectivity for AAM and UAM applications. However, the characteristic of constant LOS in the deployment of NTNs also introduces new challenges, particularly in managing interference from ground-level users \cite{cherif2020downlink, azari2017coexistence}. Concretely, the increasingly dense deployment of smart vehicles equipped with various wireless technologies on urban grounds can significantly interfere with the communication between NTNs and eVTOLs, thereby reducing the coverage reliability of UAM operation. This increased interference will degrade communication services essential for collision avoidance techniques, thereby posing potential safety risks \cite{zaid2023evtol, belmekki2022unleashing}.

Given this challenge, there is a critical need for accurate LOS models that reflect the complex urban environment where both aerial and ground-based communication systems coexist. Such a model would significantly improve the reliability of UAM communications by providing a more precise characterization of the interference and coverage behavior, and enabling a more effective deployment of both terrestrial and non-terrestrial networks \cite{azari2019cellular}. Several studies have proposed LOS probability models for aerial communication systems. The main aerial-to-ground (A2G) LOS probability model used in the literature was proposed in \cite{al2014optimal}, which was derived from an International Telecommunication Union (ITU) report in \cite{2003RECOMMENDATIONIP}. Additionally, authors in \cite{mohammed2023closed} introduced a geometric LOS probability model for UAV communications in urban environments, where they provided lower and upper bounds using the first-building-LOS probability estimate. Authors in \cite{gapeyenko2021line} utilized a Manhattan Poisson line process to model LOS probability for millimeter-wave UAV links. Furthermore, in \cite{ali2024height}, authors used airframe shadowing effects and building density variations to enhance LOS estimations for aerial networks. In \cite{ameur2024efficient}, an aerial Peer-to-Peer network is proposed for vehicular ad-hoc networks, leveraging UAVs and highlighting the importance of accurate LOS probability models. Other works highlighted the influence of road layouts and urban infrastructure on A2G communications, such as \cite{messaoudi2023uav}.


The models in the literature provide valuable insights into A2G communications, but they do not account for the impact of ground vehicles on roads. In the next section, we will show the limitation of the main model used in the literature, defined in \cite{al2014optimal}, in relation to capturing the LOS characteristics between an NTN aircraft and a ground vehicle. Thus, the contribution of this paper is to propose a model that takes into consideration the position of ground users on roads to more accurately model A2G links. Furthermore, our model parameters will be fitted using data derived from realistic simulation models and further tailored to different environmental contexts to enhance their applicability. Finally, to ensure the proposed model is applicable to real-world scenarios and environments, the proposed model will be validated using real-world urban models.

\begin{figure}[t]
    \centering
    \includegraphics[width=\columnwidth]{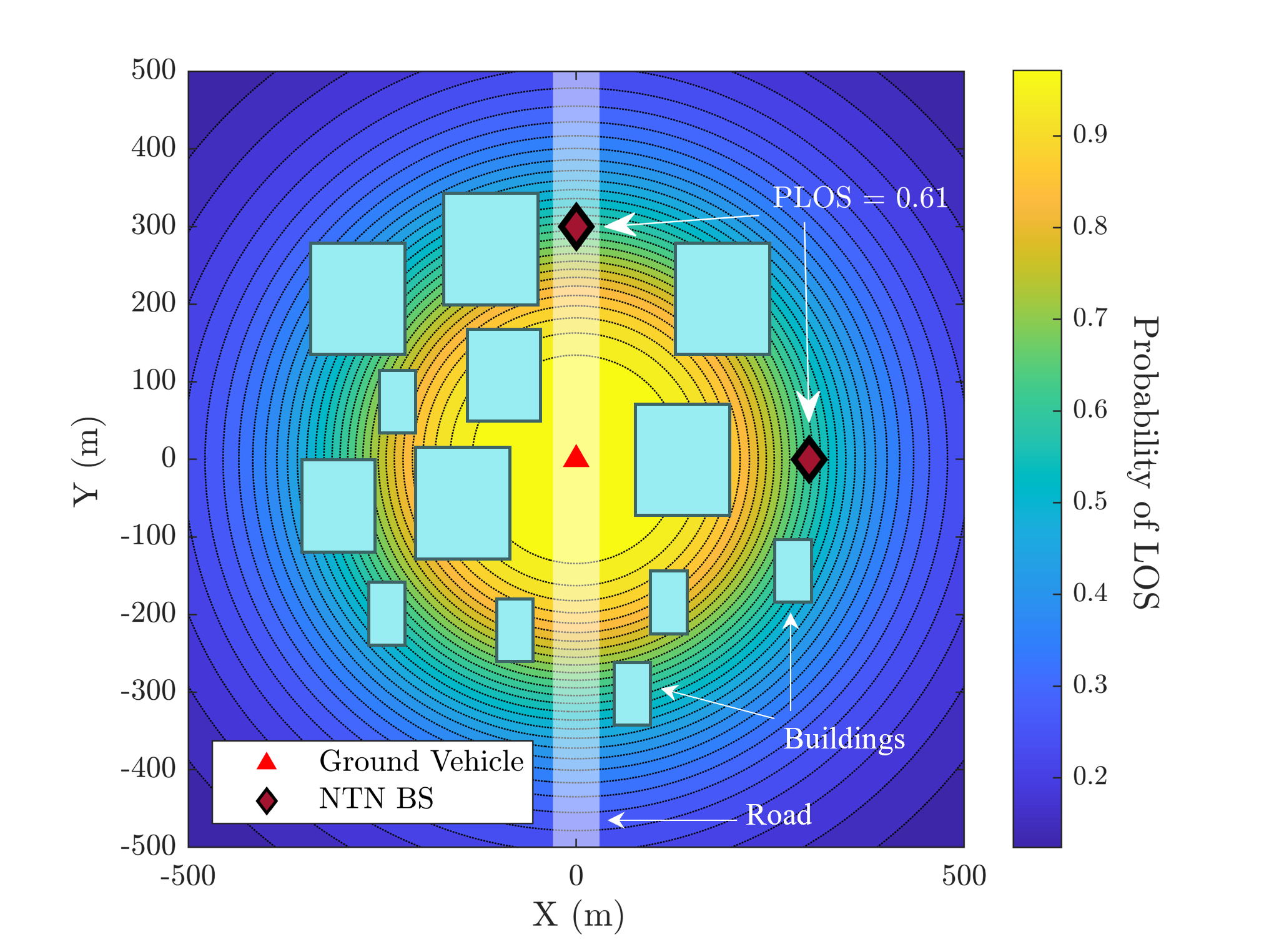}
    \caption{Traditional LOS model example showing the limitation of its use over roads.}
    \label{PLOS_Traditional}
\end{figure}

\begin{table}[]
\centering
\caption{Traditional LOS model parameters \cite{al2014modeling}.}
\resizebox{0.25\textwidth}{!}{%
\begin{tabular}{@{}ccc@{}}
\toprule
\toprule
Environment & \(a\) & \(b\) \\
\toprule
Suburban & 4.88 & 0.43 \\ 
Urban & 9.61 & 0.16 \\ 
Dense urban & 11.95 & 0.136 \\ 
High-rise urban & 27.23 & 0.08 \\ \bottomrule
\end{tabular}%
}
\label{tab:PLOS_params}
\end{table}

\section{THE TRADITIONAL LOS MODEL}
We begin by describing the traditional model from the literature, and highlighting its limitation. To achieve mathematical tractability, authors in \cite{al2014optimal} approximated the ITU model by using a modified Sigmoid function. This model depends on the elevation angle between the two users, and on specific simulated parameters that describe the density of the surrounding environment. Crucially, the typical model is independent of the ground user location in relation to its surroundings. Specifically, there is no consideration of whether the user is located on a road or elsewhere, and hence the model exhibits azimuthal symmetry. This drawback makes the typical model unable to capture the LOS probability behavior accurately between road vehicles and aerial vehicles. The traditional model (\(T\)) is defined as

\begin{equation}\label{aerialPLOS}
    P_{\textrm{LOS}}^{(T)}\left(\theta \right) = \frac{1}{1+a\textrm{exp}\left(-b\left(\theta-a\right)\right)},
\end{equation}

where \(a\) and \(b\) are parameters that reflect the density of buildings in the urban environment. The elevation angle is \(\theta=\tan^{-1}\left(h/r\right)\), where \(h\) is the altitude of the aerial node, and \(r\) is the ground distance between the source and destination. There are four environment types considered with this LOS probability model, the parameter values of these environments are given in Table \ref{tab:PLOS_params}.
The limitation of the traditional LOS model in \eqref{aerialPLOS} will become more pronounced in future urban environments with dense smart-vehicle deployment and the anticipated rise of internet-of-everything (IoE), which will increase the interference on UAM aircraft drastically \cite{zaid2023evtol}. The traditional model does not consider the streets' design; hence, every ground vehicle on any street encounters the same probability of LOS, given that the distance is equal. This limitation is demonstrated in Fig.~\ref{PLOS_Traditional}, where the probability of LOS between a ground vehicle in the origin with an NTN flying at a specific height is plotted as a heatmap.  We consider two NTNs flying with equal distance to the vehicle, however one of the NTNs is flying above the road, while the other is above buildings. The model predicts that both NTNs would have the same probability of LOS, regardless of the road. The limitation in this model is that the roads' airspace is generally empty, and thus, most vehicles would interfere on a UAM aircraft with a high probability of LOS. In the subsequent part, we propose an extended LOS model that addresses this limitation.

\begin{figure}[t]
    \centering
    \includegraphics[width=0.3\textwidth]{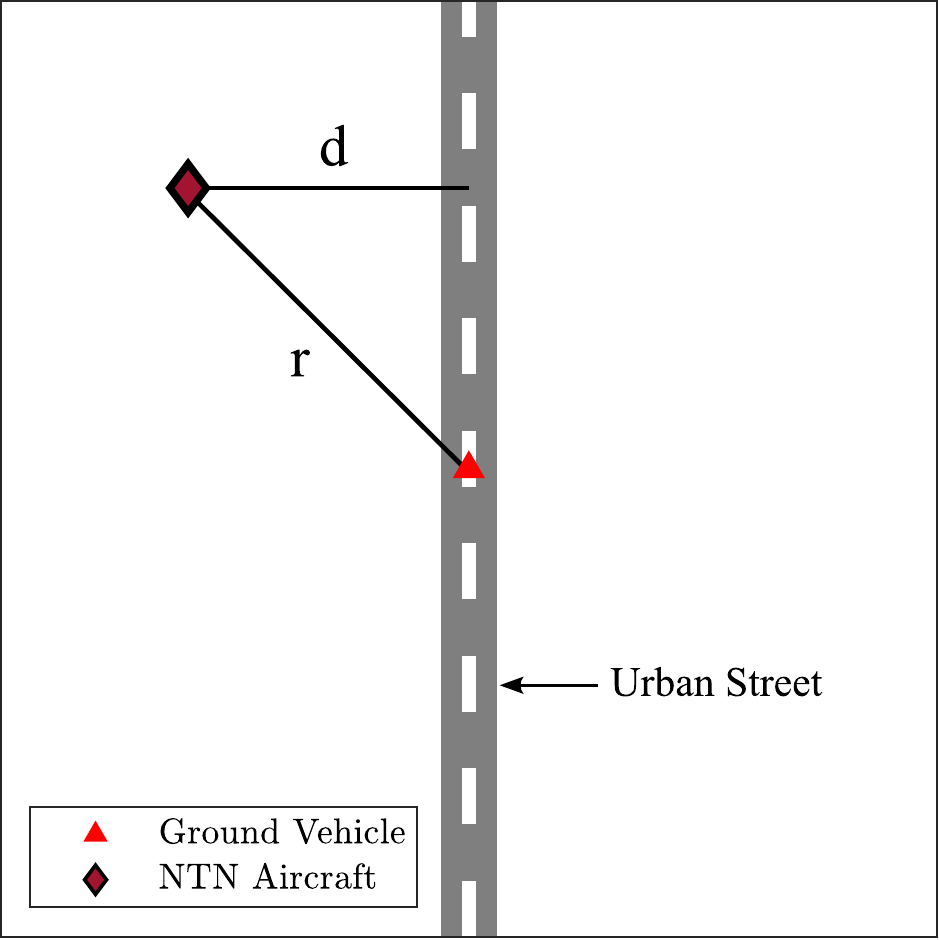}
    \caption{The perpendicular distance \(d\) considered for the extended LOS probability model. \(r\) is the ground distance.}
    \label{PLOS_Novel_parameters}
\end{figure}

\section{A ROAD-AWARE LOS MODEL}
To design a model that accounts for the unique interactions between NTNs and ground vehicles, we introduce a street-aware modification to the traditional model, which incorporates a new function that adjusts the LOS probability based on the perpendicular distance \(d\) between the NTN and the road, shown in Fig.~\ref{PLOS_Novel_parameters}. Now, instead of treating every location uniformly, our extended model blends two probabilities: (i) the traditional environment-based LOS probability and (ii) a street-dependent LOS probability that reflects the reduced obstructions over the road. This adjustment ensures that the LOS probability is higher when the NTN is directly above the road, and smoothly transitions to the traditional environment-based LOS probability as the distance from the road increases. Our proposed extended model (\(E\)) is given as
\begin{equation}
    \begin{split}
    P_{\text{LOS}}^{(E)}(\theta) & =
    \exp(-cd^2) \times 
    \underbrace{\frac{1}{1 + a_1\exp(-b_1(\theta-a_1))}}_{\text{function of street}}\\ 
    &\!\!\!\! + (1 - \exp(-cd^2)) \times \underbrace{\frac{1}{1 + a\exp(-b(\theta-a))}}_{\text{function of environment}},
    \end{split}
\end{equation}
where we have defined three new parameters, \(c\), \(a_1\), and  \(b_1\). The parameter \(c\) is influenced by and indicative of the street width, where \(a_1\) and \(b_1\) are parameters that describe the street density in relation to vehicle congestion. The width of a road directly affects the LOS probability, since wider roads generally provide a larger obstruction-free region, reducing the probability of buildings blocking the LOS, this is captured by the parameter \(c\). As in the traditional model, \(\theta=\tan^{-1}\left(h/r\right)\) is the elevation angle. The term \(\exp(-cd^2)\) serves as a spatial weighting function, ensuring that the transition from the road-dominated LOS probability to the environment-based LOS probability is smooth. The choice of an exponential function is motivated by the need for a gradual shift in the influence of roads as the distance increases. The parameter \(c\) determines the rate at which this transition occurs. Furthermore, when the NTN is directly above the road (\(d=0\)), the weighting function approaches \(1\), meaning the street-based function dominates. As \(d\) increases, the model transitions back to the traditional LOS model, i.e.
\begin{equation}
    \begin{split}
    \Lim{d \rightarrow \infty} P_{\text{LOS}}^{(E)}(\theta) & =
    P_{\text{LOS}}^{(T)}(\theta).
    \end{split}
\end{equation}

\begin{figure}[t]
    \centering
    \includegraphics[width=\columnwidth]{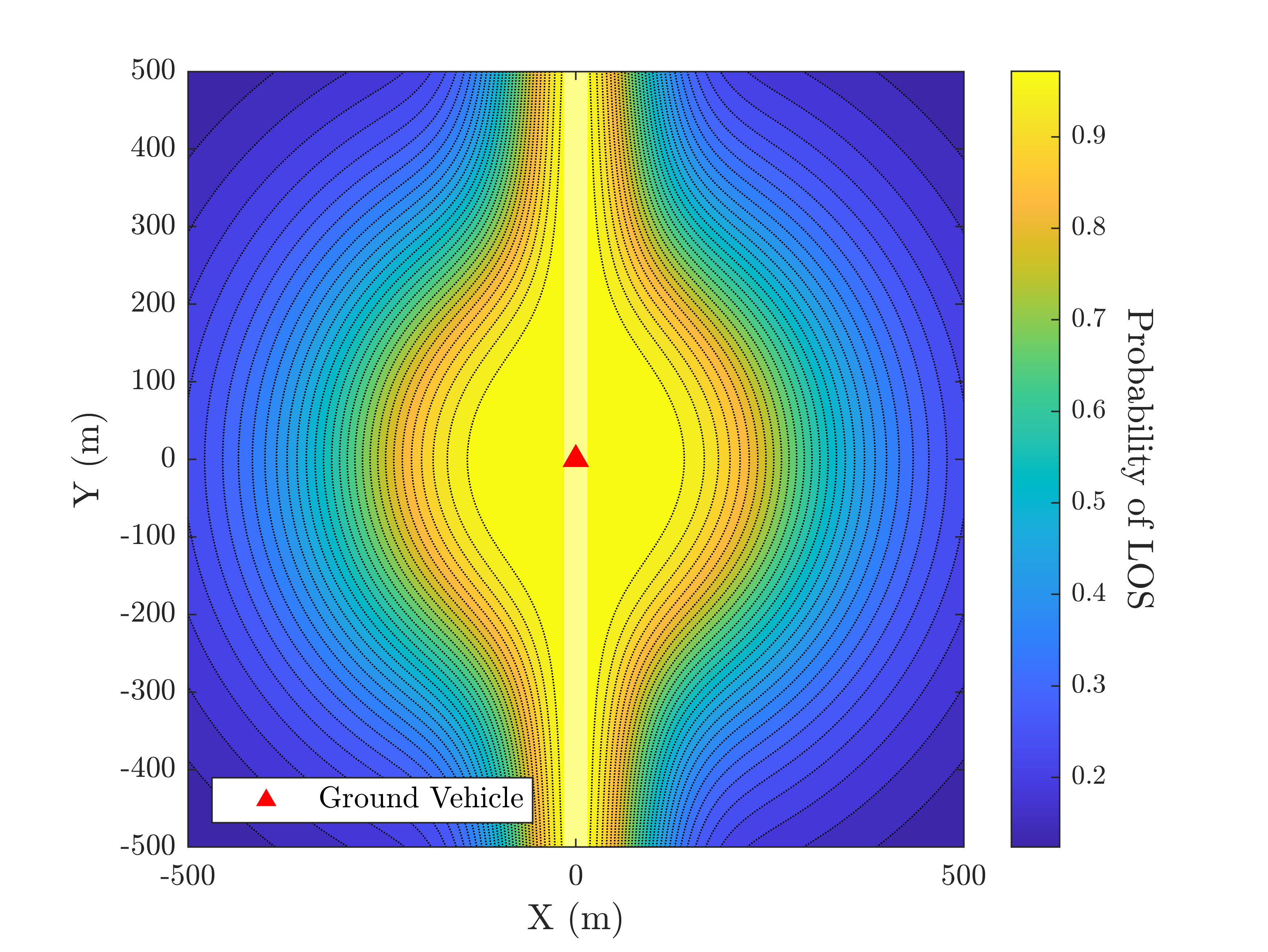}
    \caption{Extended LOS probability model heatmap curve using arbitrary parameters.}
    \label{PLOS_Novel_example}
\end{figure}

Fig.~\ref{PLOS_Novel_example} shows the proposed model plotted using arbitrary parameters for illustration, where we notice the model is clearly able to capture the behavior of the LOS probability above the street. Concretely, the model predicts a high probability of LOS along the street, while preserving the same azimuthal symmetry behavior for aircraft flying away from the street. This outcome aligns well with our expectations.

\begin{figure}[t]
    \centering
    \includegraphics[width=\columnwidth]{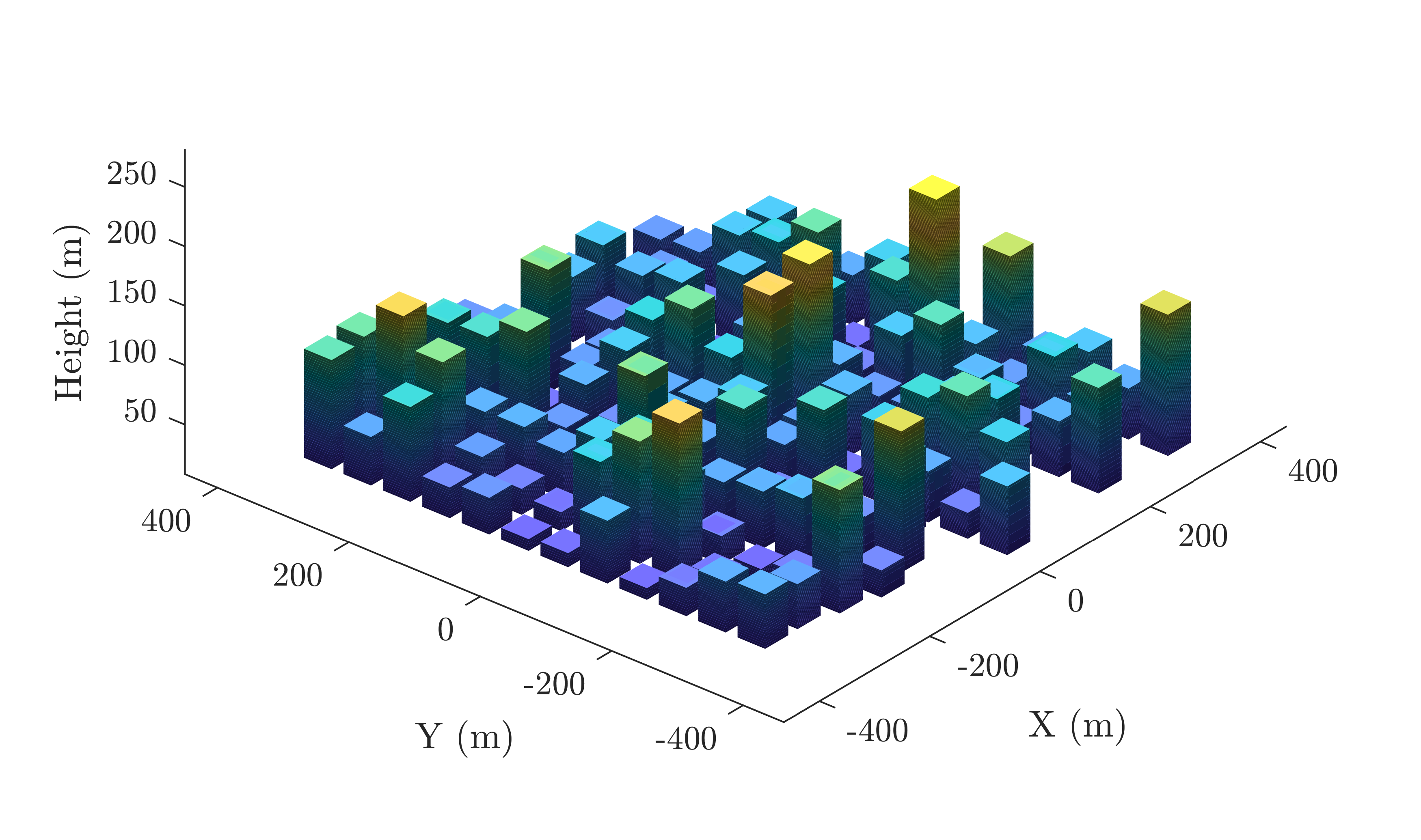}
    \caption{Instance of a simulated high-rise urban environment used for parameter fitting.}
    \label{City_highrise}
\end{figure}

\begin{table}[t]
\centering
\caption{LOS probability environment simulation parameters \cite{al2014modeling}. \(\alpha_0\) represents the  ratio of built-up land area to the total area (dimensionless), \(\beta_0\) represents the mean number of buildings per unit area (buildings/\(\text{km}^2\)), and \(\gamma_0\) is a scale parameter that describes the buildings heights according to a Rayleigh distribution.}
\resizebox{0.3\textwidth}{!}{%
\begin{tabular}{@{}cccc@{}}
\toprule
\toprule
Environment & $\alpha_0$ & $\beta_0$ & $\gamma_0$ \\
\toprule
Suburban             & 0.1                  & 750                  & 8                    \\
Urban                & 0.3                  & 500                  & 15                   \\ 
Dense Urban          & 0.5                  & 300                  & 20                   \\ 
High-rise Urban       & 0.5                  & 300                  & 50                   \\ \bottomrule
\end{tabular}%
}
\label{tab:PLOS_building_params}
\end{table}

\begin{table}[t]
\centering
\caption{Extended LOS probability model fitted parameters.}
\resizebox{\columnwidth}{!}{%
\begin{tabular}{@{}cccccc@{}}
\toprule
\toprule
Environment & \(a\) & \(b\) & \(c\) & \(a_1\) & \(b_1\) \\ \midrule
Urban & 9.61 & 0.16 & 0.0014 & 0.26 & 0.05 \\ 
Dense urban & 11.95 & 0.136 & 0.0019 & 2.63 & 0.11 \\ 
High-rise urban & 27.23 & 0.08 & 0.0044 & 6.84 & 0.18 \\ \bottomrule
\end{tabular}%
}
\label{tab:PLOS_Novel_params}
\end{table}

\begin{figure*}[t]
    \centering
    \begin{subfigure}[b]{0.32\textwidth}
        \centering
        \includegraphics[width=\textwidth]{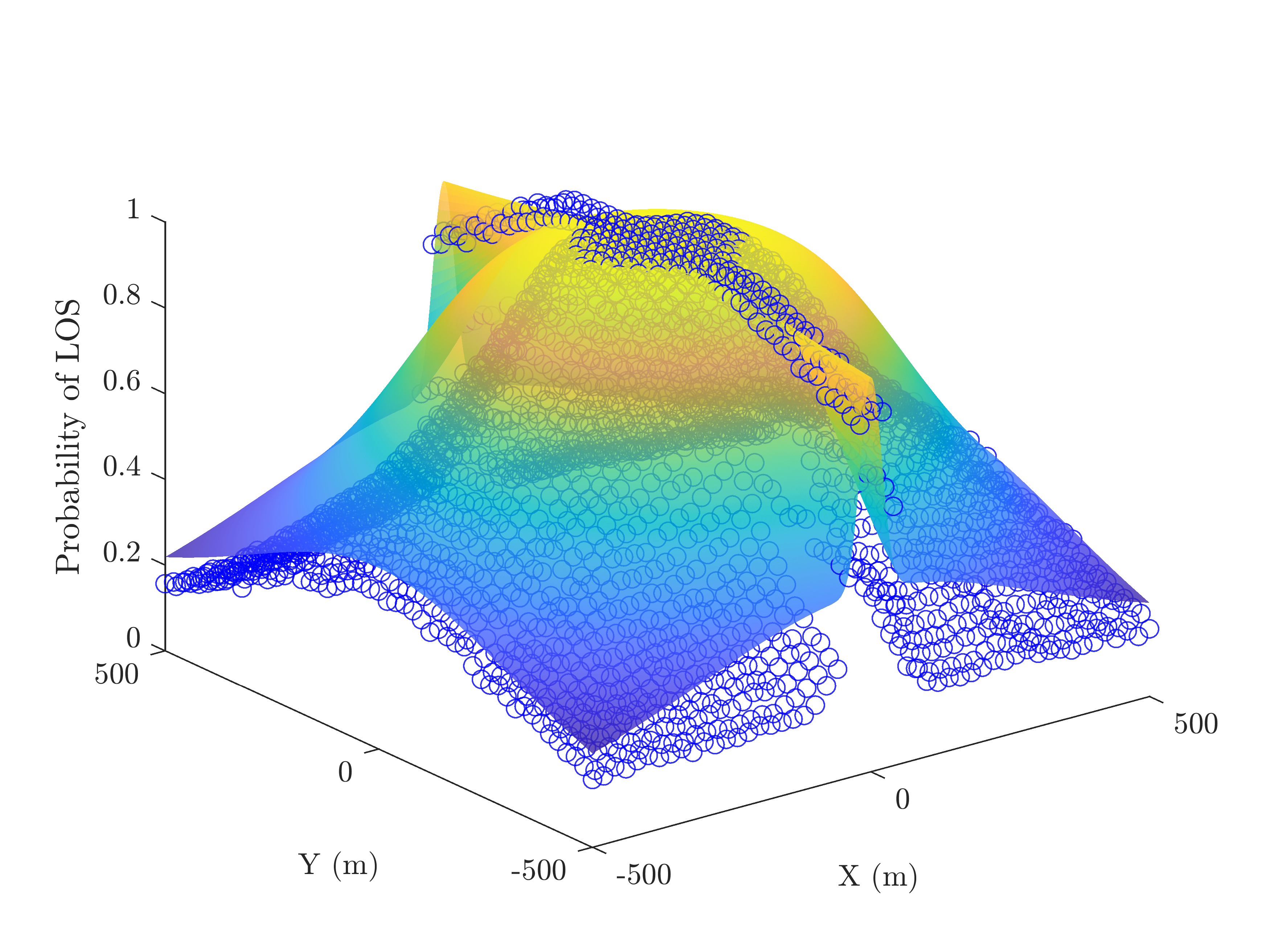}
        \caption{Simulation - Urban.}
        \label{fig:sub1}
    \end{subfigure}
    \hfill 
    \begin{subfigure}[b]{0.32\textwidth}
        \centering
        \includegraphics[width=\textwidth]{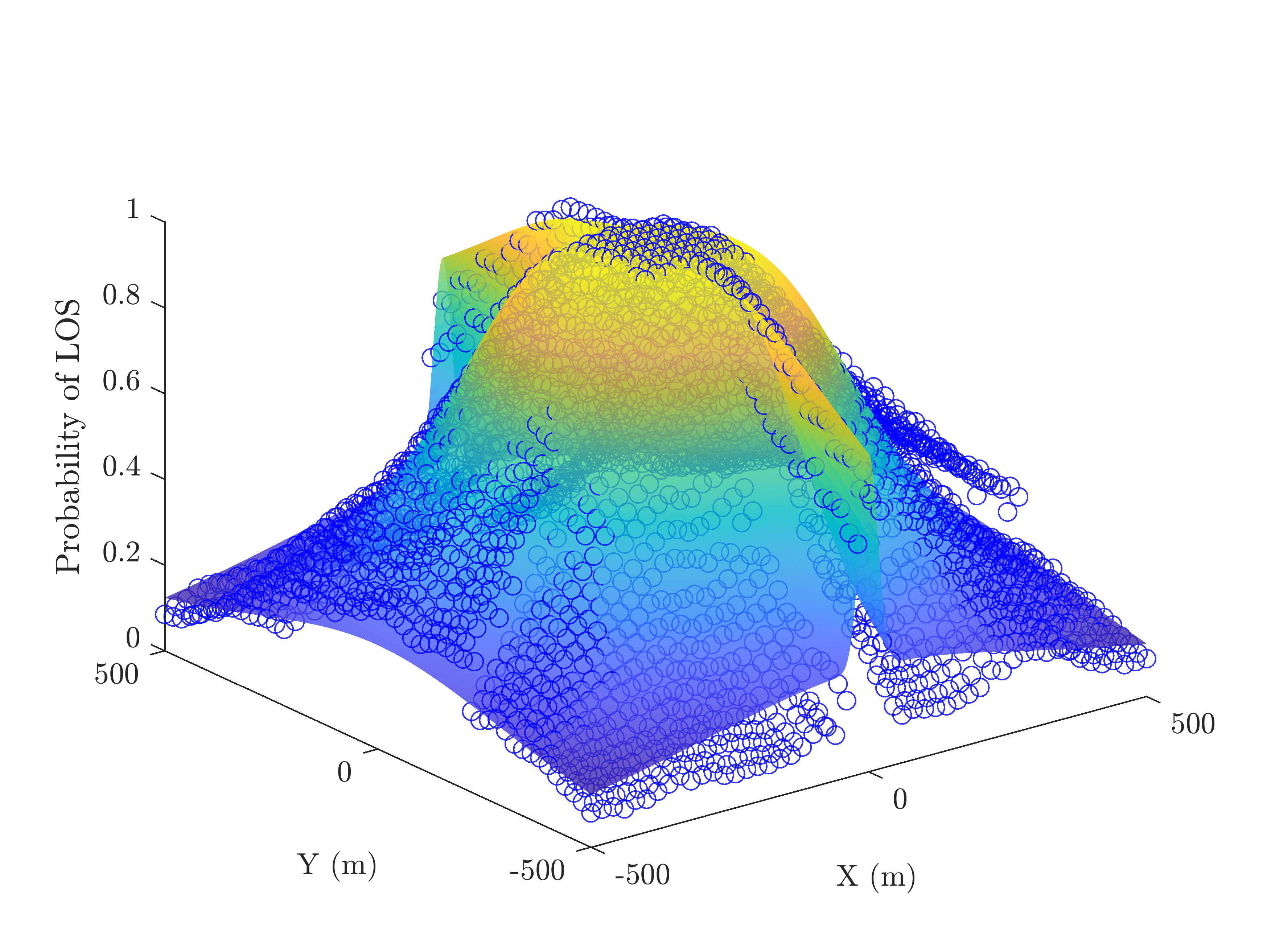}
        \caption{Simulation - Dense Urban.}
        \label{fig:sub2}
    \end{subfigure}
    \hfill 
    \begin{subfigure}[b]{0.32\textwidth}
        \centering
        \includegraphics[width=\textwidth]{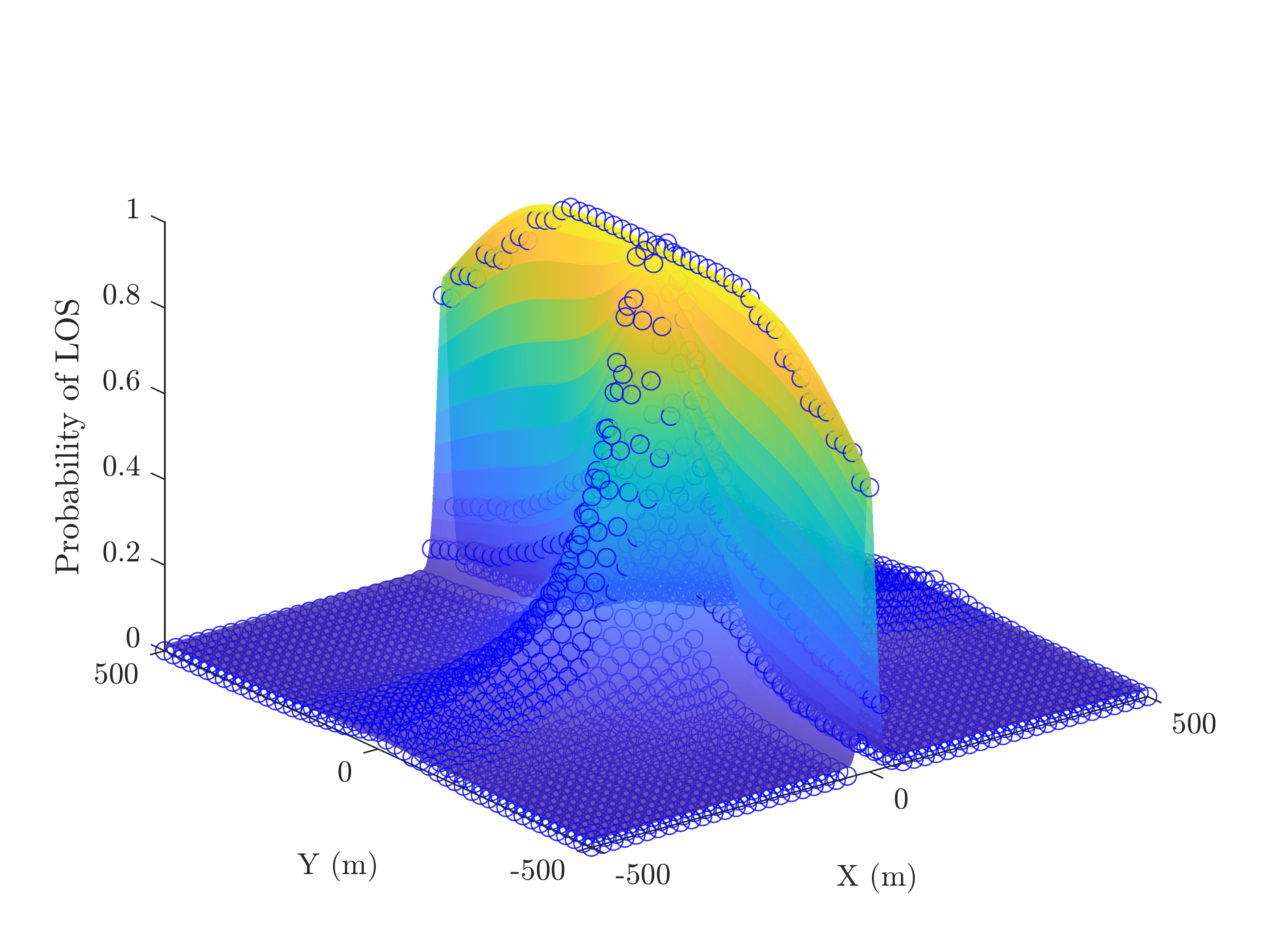}
        \caption{Simulation - High-rise.}
        \label{fig:sub3}
    \end{subfigure}
    \vspace{0.5cm} 
    \begin{subfigure}[b]{0.33\textwidth}
        \centering
        \includegraphics[width=\textwidth]{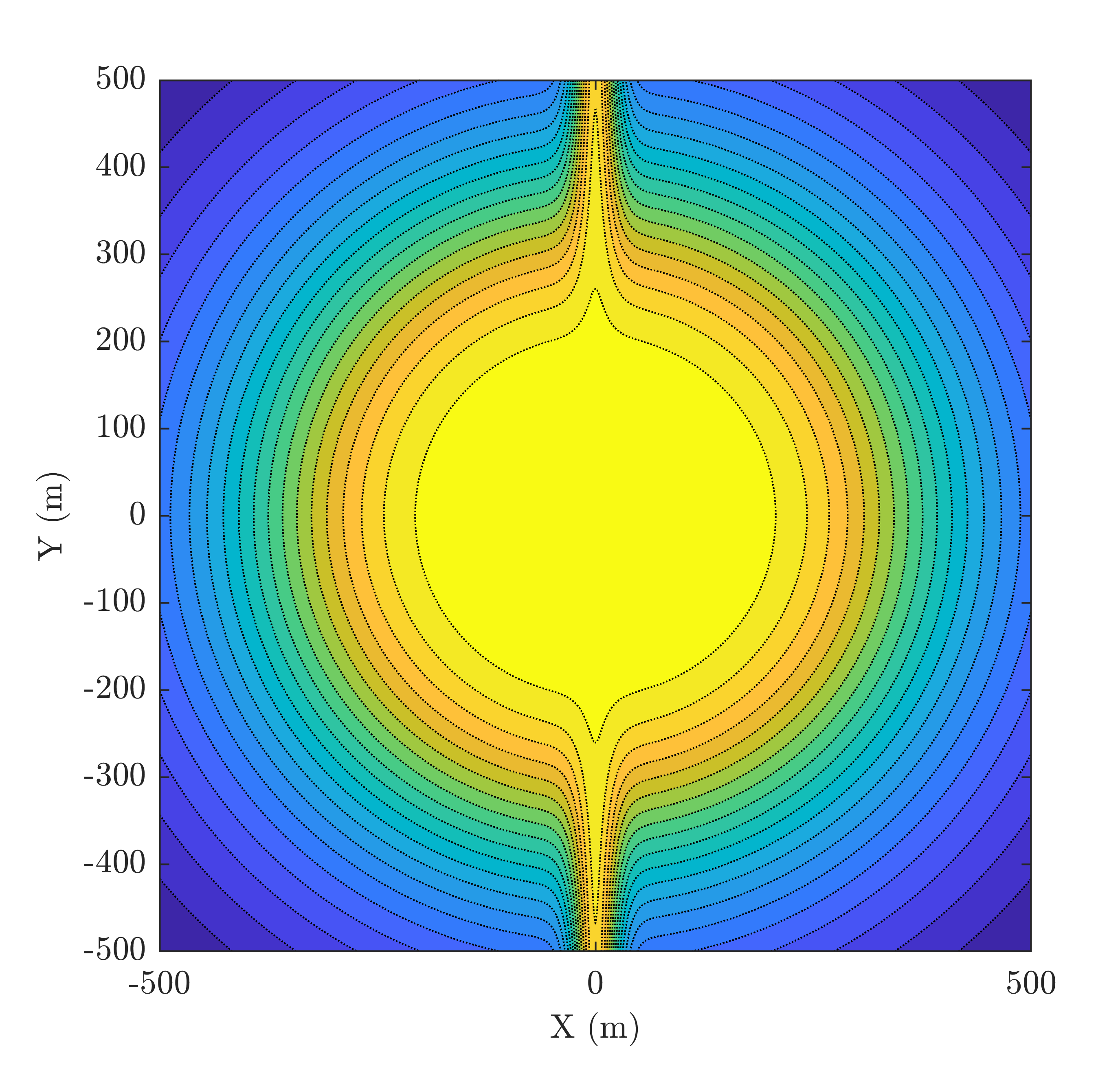}
        \caption{Fitted model - Urban.}
        \label{fig:sub4}
    \end{subfigure}
    \hfill 
    \begin{subfigure}[b]{0.32\textwidth}
        \centering
        \includegraphics[width=\textwidth]{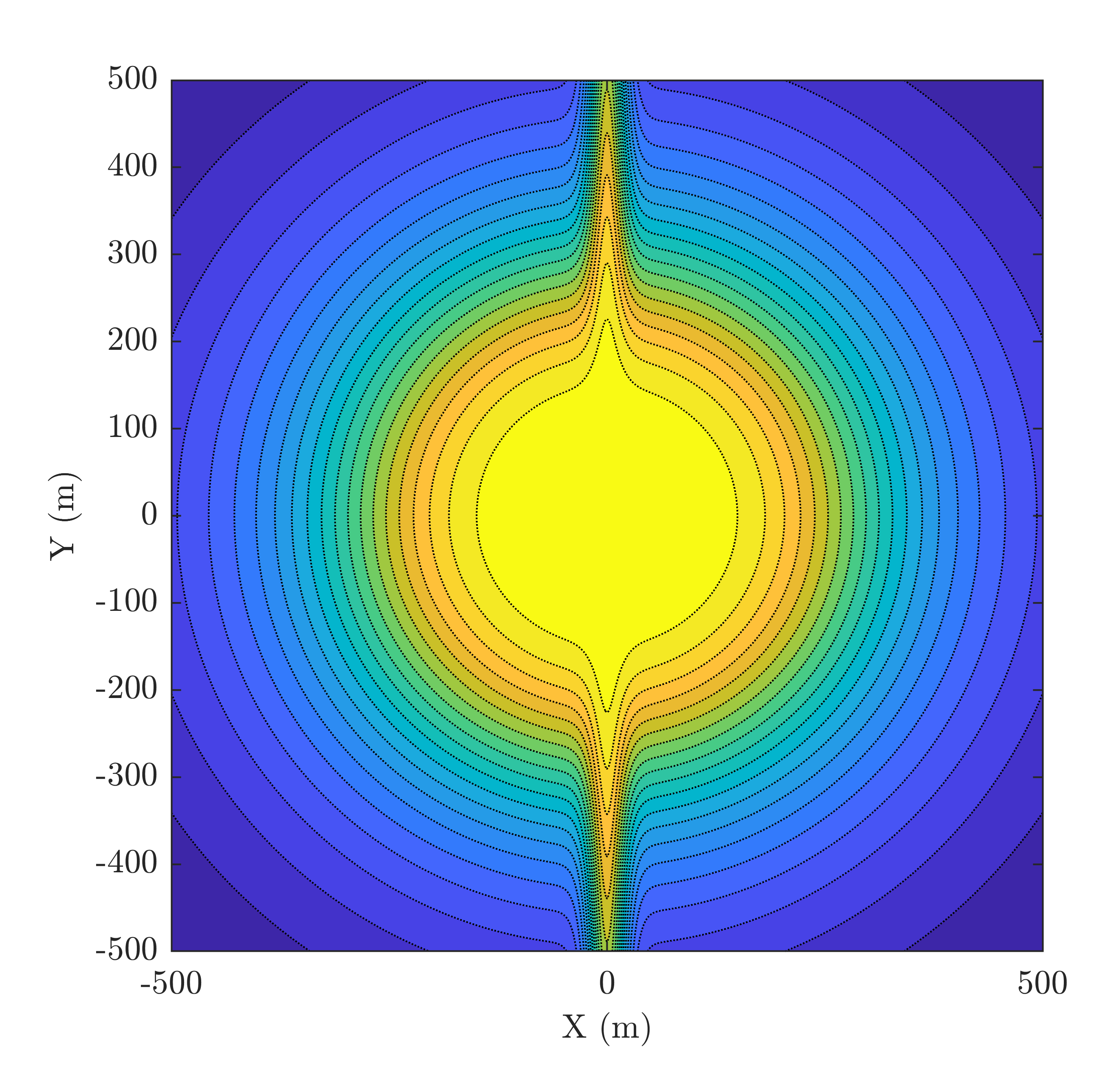}
        \caption{Fitted model - Dense Urban.}
        \label{fig:sub5}
    \end{subfigure}
    \hfill 
    \begin{subfigure}[b]{0.32\textwidth}
        \centering
        \includegraphics[width=\textwidth]{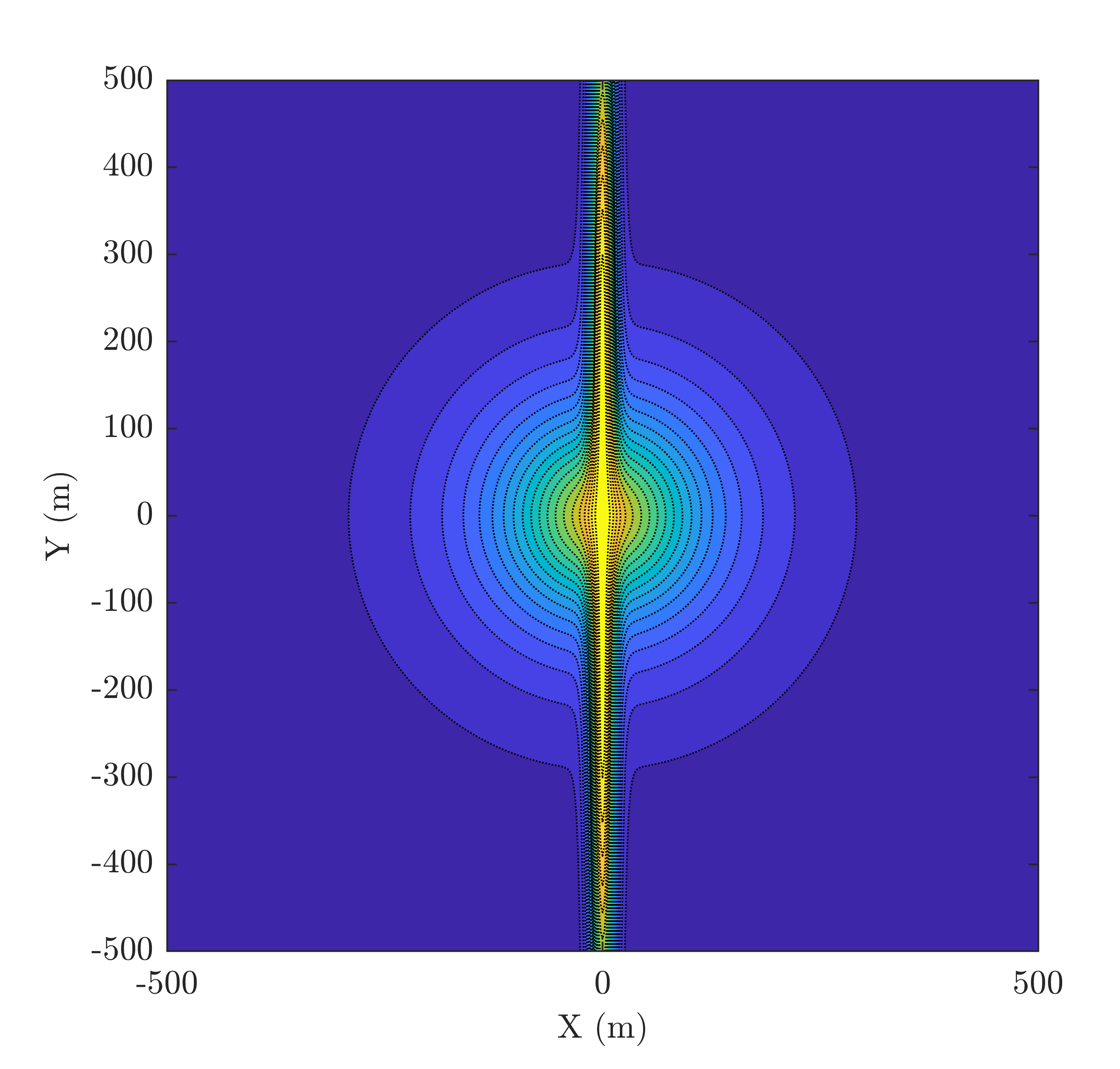}
        \caption{Fitted model - High-rise.}
        \label{fig:sub6}
    \end{subfigure}
    \caption{Fitted simulation data in (a-c), and plotted model with the fitted parameters in (d-f).}
    \label{fig:Fitted_model_figures}
\end{figure*}

\section{PARAMETER FITTING}
To use our proposed model in real-case scenarios, the new parameters introduced, that is, \(c\), \(a_1\), and \(b_1\), must be defined well and according to realistic assumptions. To fit the parameters, we conduct a simulation using MATLAB according to the following steps. First, we build an environment consisting of buildings and a street with vehicles. The buildings will be distributed according to the parameters in Table~\ref{tab:PLOS_building_params}. These environmental parameters for suburban, urban, dense urban, and high-rise urban settings are derived from the ITU recommendations on urban morphology \cite{2003RECOMMENDATIONIP}. We utilize the building density and height distributions for each environment, which are based on empirical data. Using these parameters and assuming square-shaped buildings, we can determine both the building dimensions and the spacing between them. Specifically, the (dimension, separation) values for each environment type are as follows: Suburban: (\(12~\)m, \(26~\)m), Urban: (\(25~\)m, \(21~\)m), Dense/High-rise Urban: (\(41~\)m, \(19~\)m). An instance of this urban simulation for a high-rise urban environment is shown in Fig.~\ref{City_highrise}. Second, using this simulated city, we run a ray-tracing simulation that calculates the existence of a LOS link between a ground vehicle at the origin with an NTN over $\mathbb{R}^2$. This simulation is  repeated \(800\) times and averaged over the buildings' heights. Finally, using this simulated LOS probability data, we fit our proposed model to this data by using the least squares regression method for parameter estimation. This approach aligns with prior studies on environmental parameterization in UAV communication networks, such as in \cite{bouzid20235g}.

The fitted parameters obtained by using this method are given in Table~\ref{tab:PLOS_Novel_params}. Interestingly, the sparse and low density of buildings in suburban environments renders the presence of the vehicle on a street irrelevant to the probability of LOS. Consequently, our analysis determines that the traditional LOS model is sufficient for low-density environments. The simulated probabilities and the fitted curves are shown in Fig.~\ref{fig:Fitted_model_figures}

\begin{figure*}[t]
    \centering
    \begin{subfigure}[b]{0.3\textwidth}
        \centering
        \includegraphics[width=\textwidth]{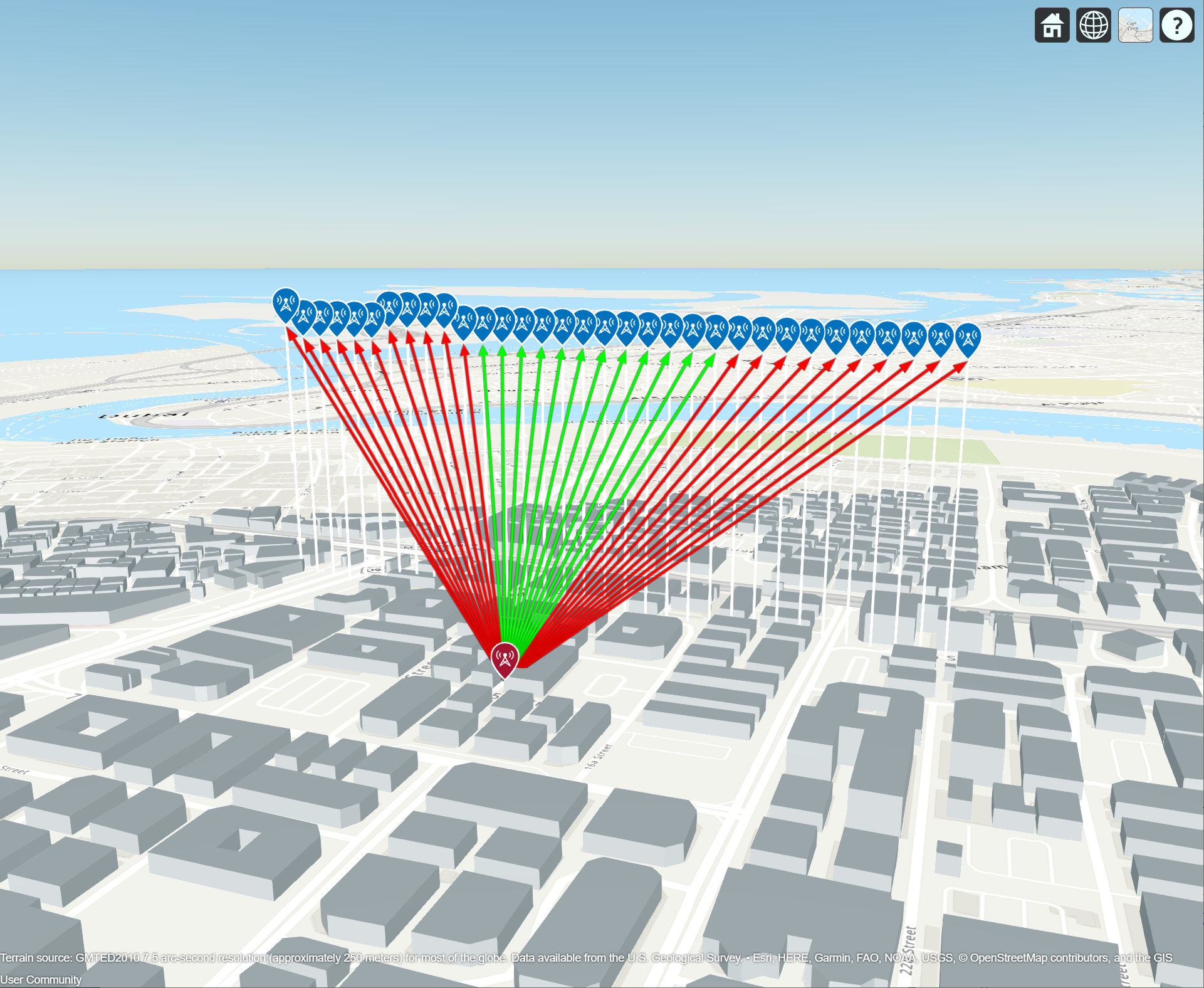}
        \caption{Simulation - Urban (Dubai).}
        \label{fig:Dubai_3D}
    \end{subfigure}
    \hfill 
    \begin{subfigure}[b]{0.3\textwidth}
        \centering
        \includegraphics[width=\textwidth]{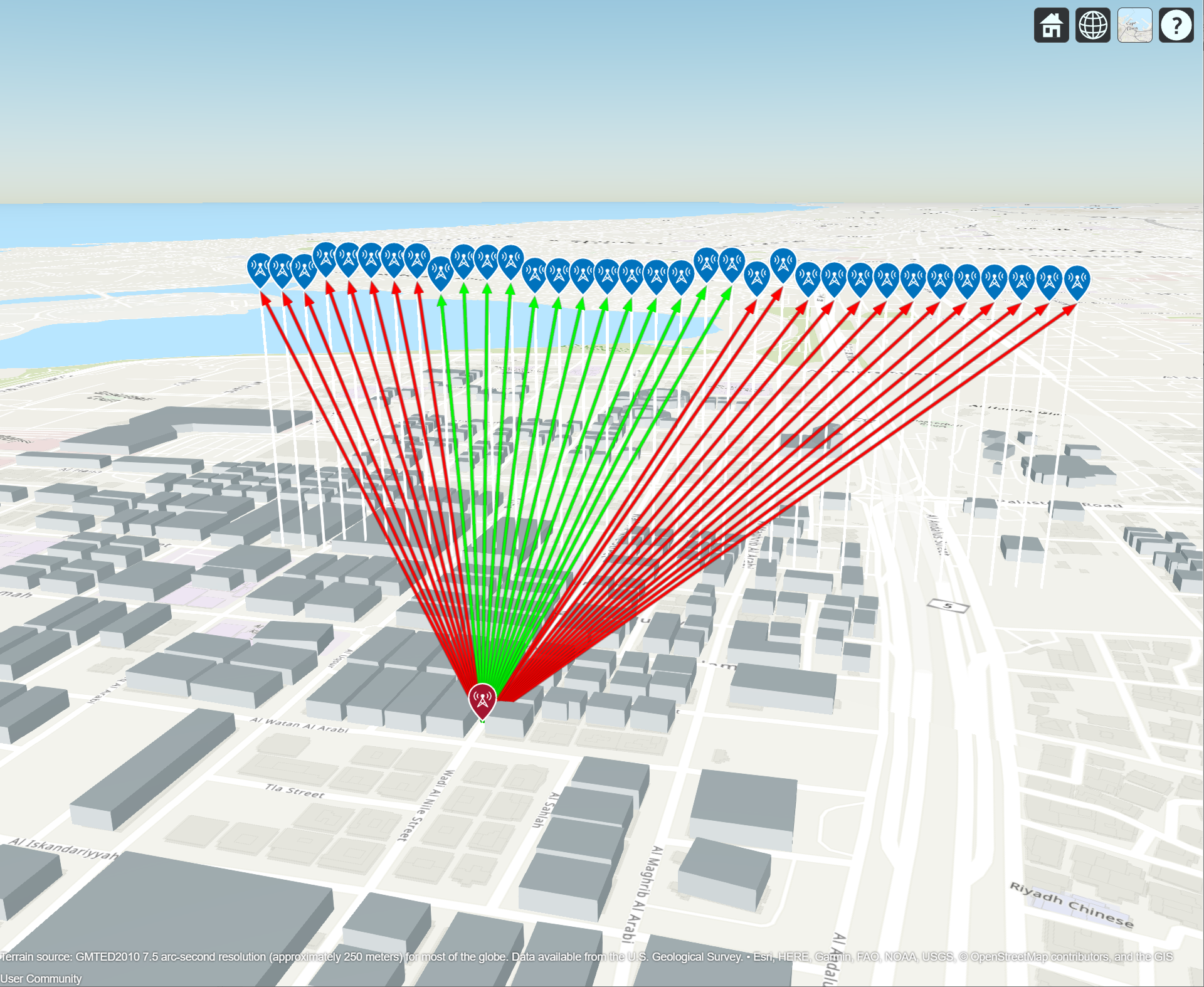}
        \caption{Simulation - Dense Urban (Jeddah).}
        \label{fig:Jeddah_3D}
    \end{subfigure}
    \hfill 
    \begin{subfigure}[b]{0.3\textwidth}
        \centering
        \includegraphics[width=\textwidth]{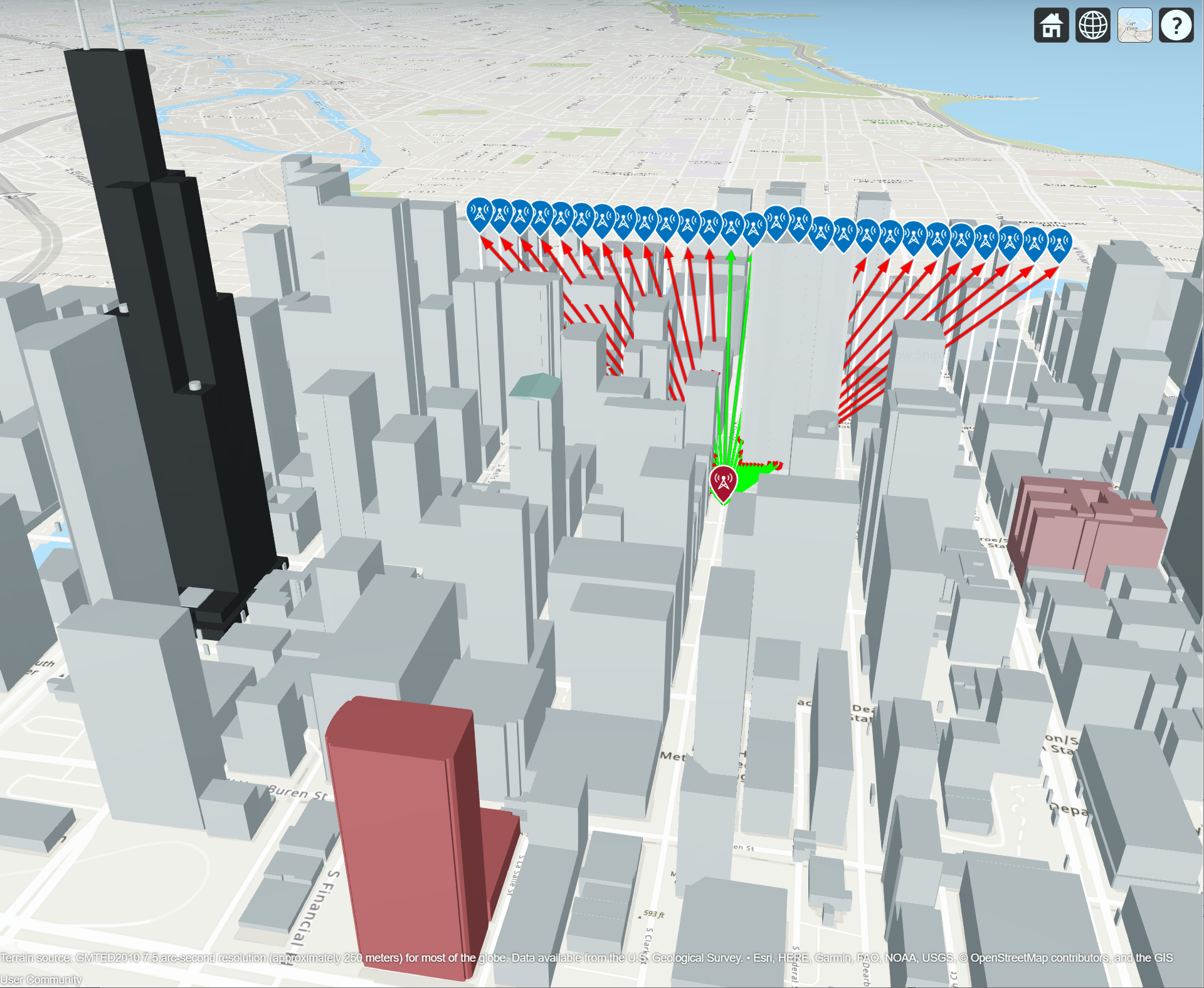}
        \caption{Simulation - High-rise (Chicago).}
        \label{fig:Chicago_3D}
    \end{subfigure}
    \vspace{0.5cm} 
    \begin{subfigure}[b]{0.32\textwidth}
        \centering
        \includegraphics[width=\textwidth]{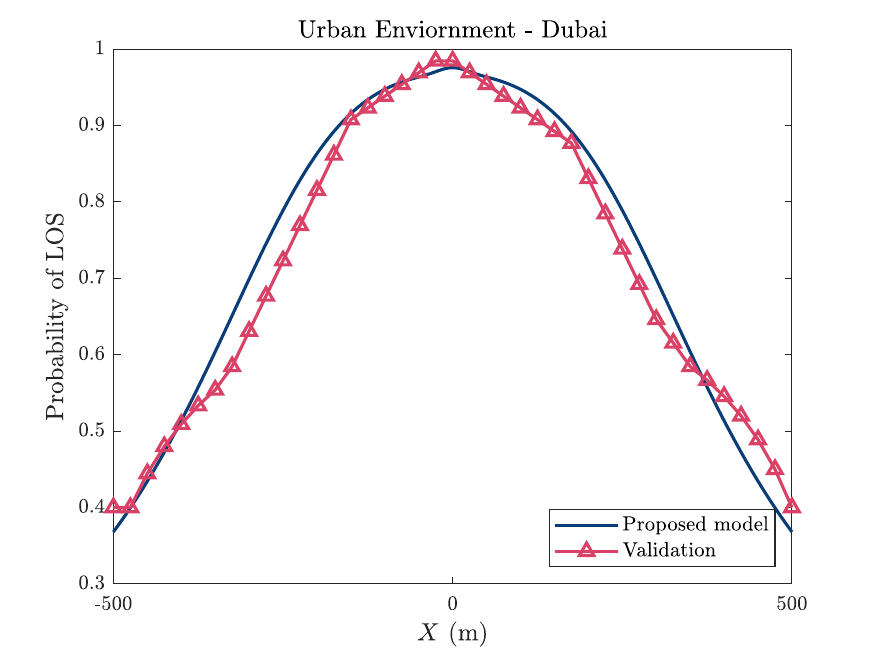}
        \caption{Validation - Urban (Dubai).}
        \label{fig:Validation_dubai}
    \end{subfigure}
    \hfill 
    \begin{subfigure}[b]{0.32\textwidth}
        \centering
        \includegraphics[width=\textwidth]{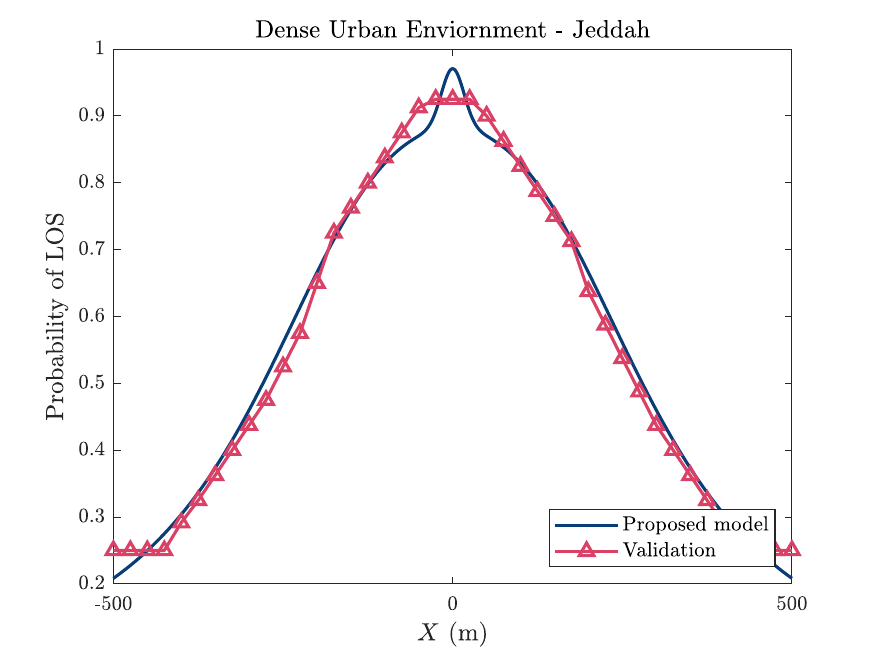}
        \caption{Validation - Dense Urban (Jeddah).}
        \label{fig:Validation_jeddah}
    \end{subfigure}
    \hfill 
    \begin{subfigure}[b]{0.32\textwidth}
        \centering
        \includegraphics[width=\textwidth]{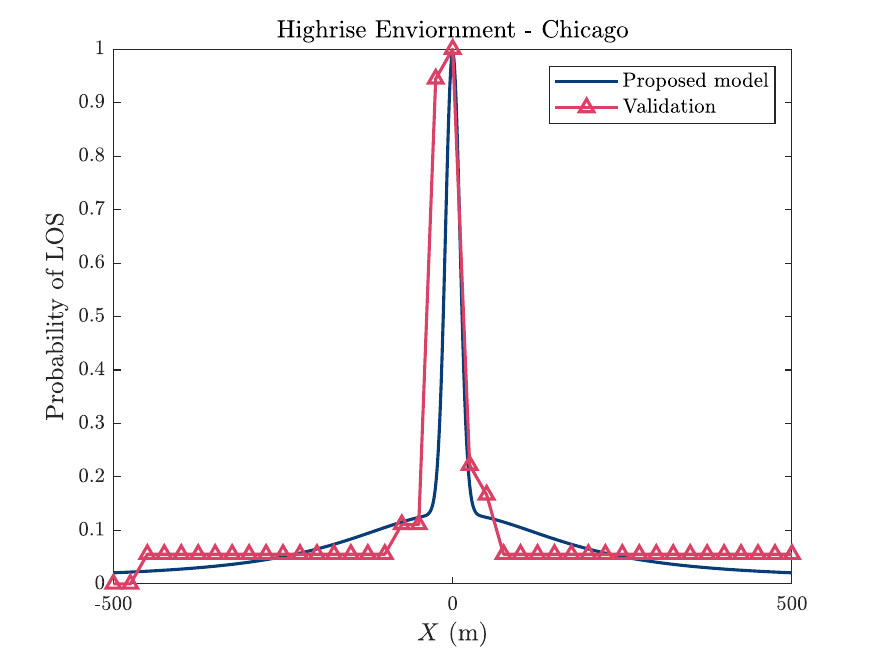}
        \caption{Validation - High-rise (Chicago).}
        \label{fig:Validation_chicago}
    \end{subfigure}
    \caption{Ray-tracing results from OSM in (a-c), and validation results in (d-f).}
    \label{fig:Validated_model_figures}
\end{figure*}

\section{MODEL VALIDATION}
With the fitted model parameters now determined, the next critical step is to validate them in real-world urban environments. To achieve this, real-world three-dimensional (3D) building models obtained from OpenStreetMap (OSM) were used \cite{OpenStreetMap}. A segment of Dubai, United Arab Emirates, was selected to represent an urban environment, a section of Jeddah, Saudi Arabia, was chosen to depict a dense urban setting, and an area of downtown Chicago, United States of America, was picked to illustrate a high-rise environment. The validation is carried out by fixing a vehicle on the ground on a street, and simulating an NTN flying on a path perpendicular to the street. Then, a ray-tracing algorithm is run to determine if the link at this specific location is in LOS or not. Figs.~\ref{fig:Dubai_3D}, \ref{fig:Jeddah_3D}, and \ref{fig:Chicago_3D} demonstrate the 3D OSM models and shows instances of the ray-tracing algorithm and the LOS results. Figs.~\ref{fig:Validation_dubai}, \ref{fig:Validation_jeddah}, and \ref{fig:Validation_chicago} show the validation curves. The OSM simulation figures show red arrows, which represent a NLOS link, and green arrows, which represent a LOS link. Next, we repeat the computation at ten different locations along the road and average the resulting LOS probabilities. The validation curves show that our model closely matches the simulation data, with root mean square errors of \(0.366\), \(0.228\), and \(0.1216\) for urban, dense urban, and high-rise environments, respectively.

\begin{figure*}[t]
    \centering
    \begin{subfigure}[b]{0.329\textwidth}
        \centering
        \includegraphics[width=\columnwidth]{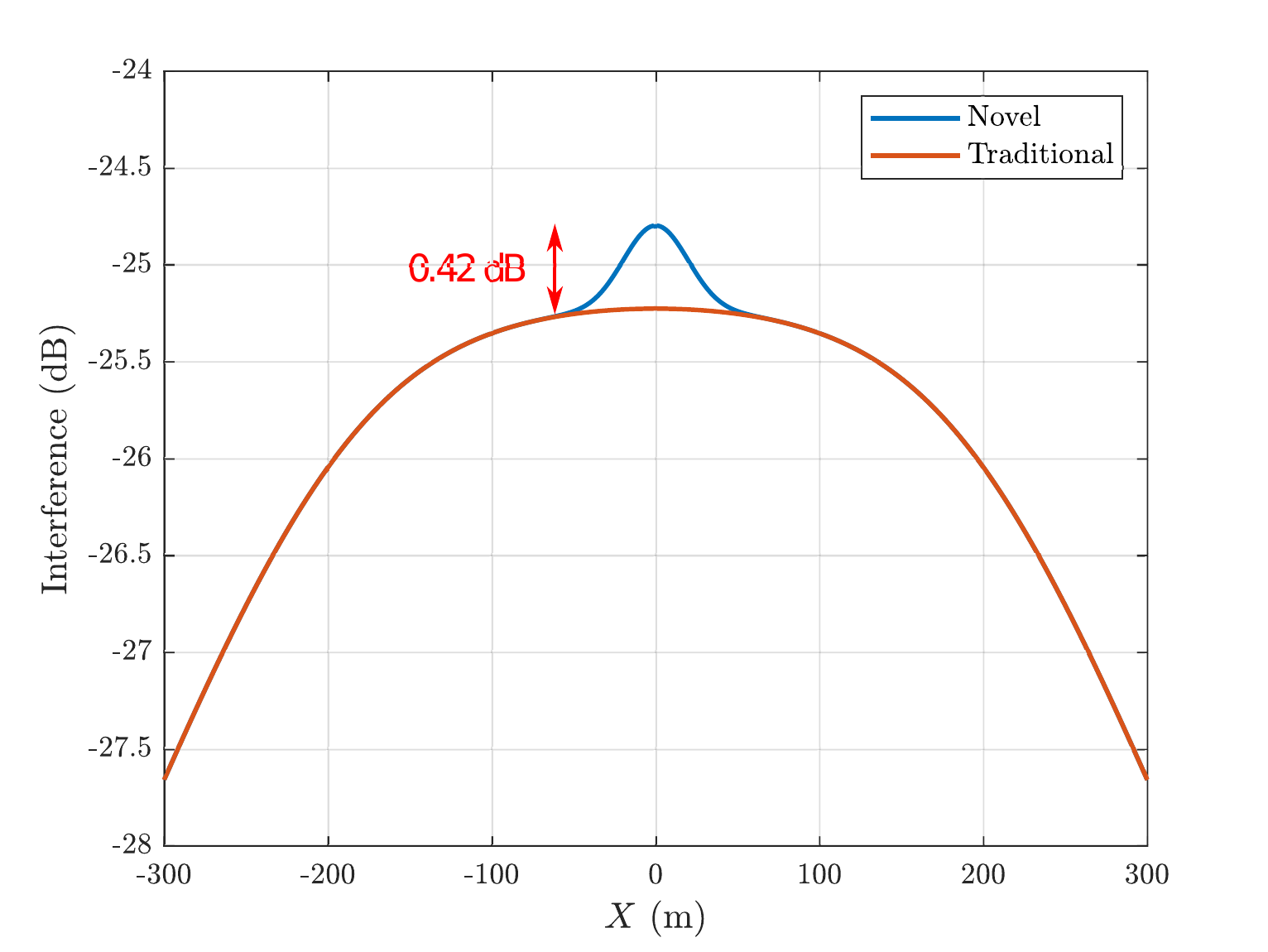}
        \caption{Urban.}
        \label{fig:Interference_difference_urban}
    \end{subfigure}
    \begin{subfigure}[b]{0.329\textwidth}
        \centering
        \includegraphics[width=\columnwidth]{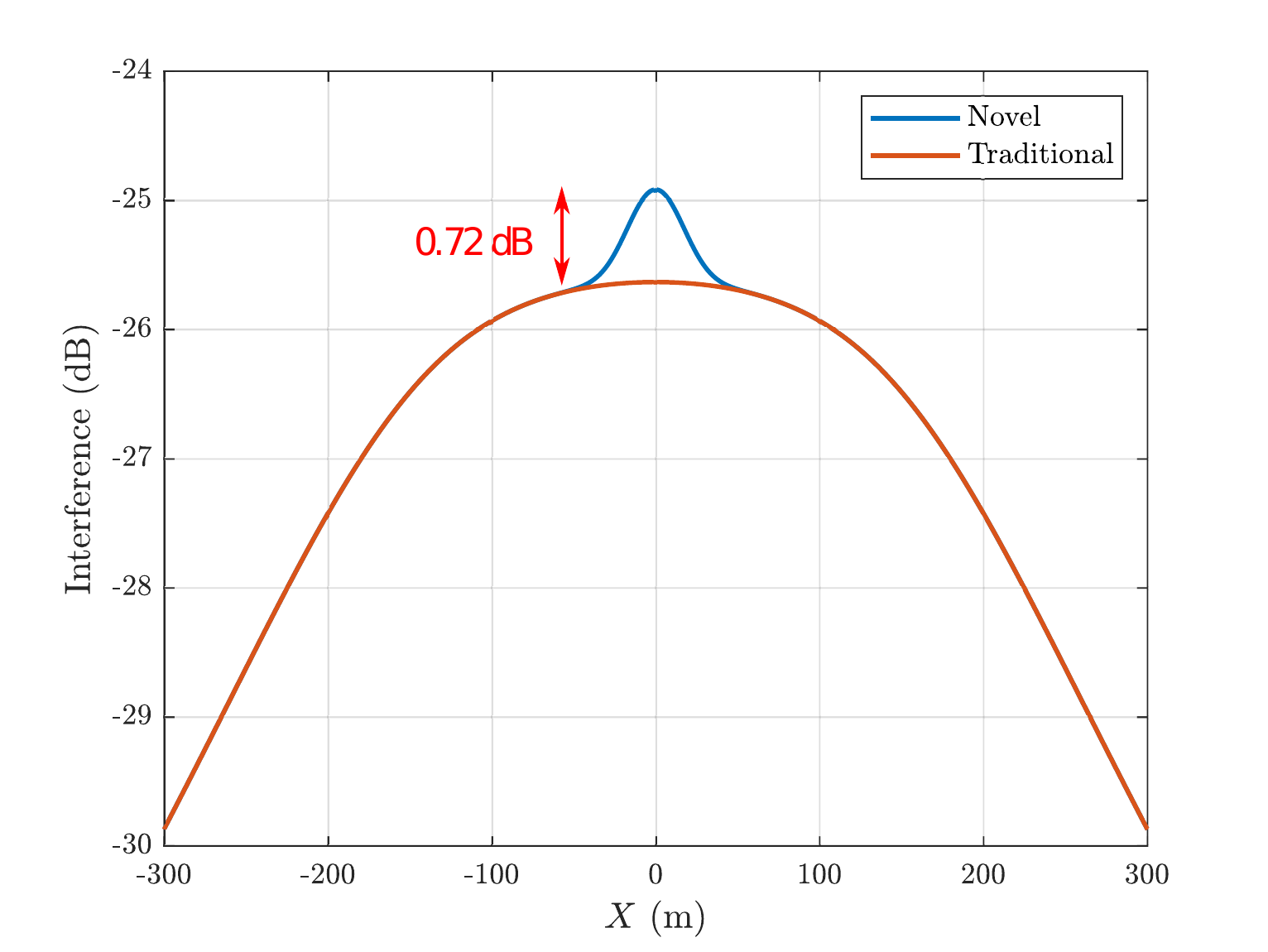}
        \caption{Dense Urban.}
        \label{fig:Interference_difference_denseurban}
    \end{subfigure}
    \begin{subfigure}[b]{0.329\textwidth}
        \centering
        \includegraphics[width=\columnwidth]{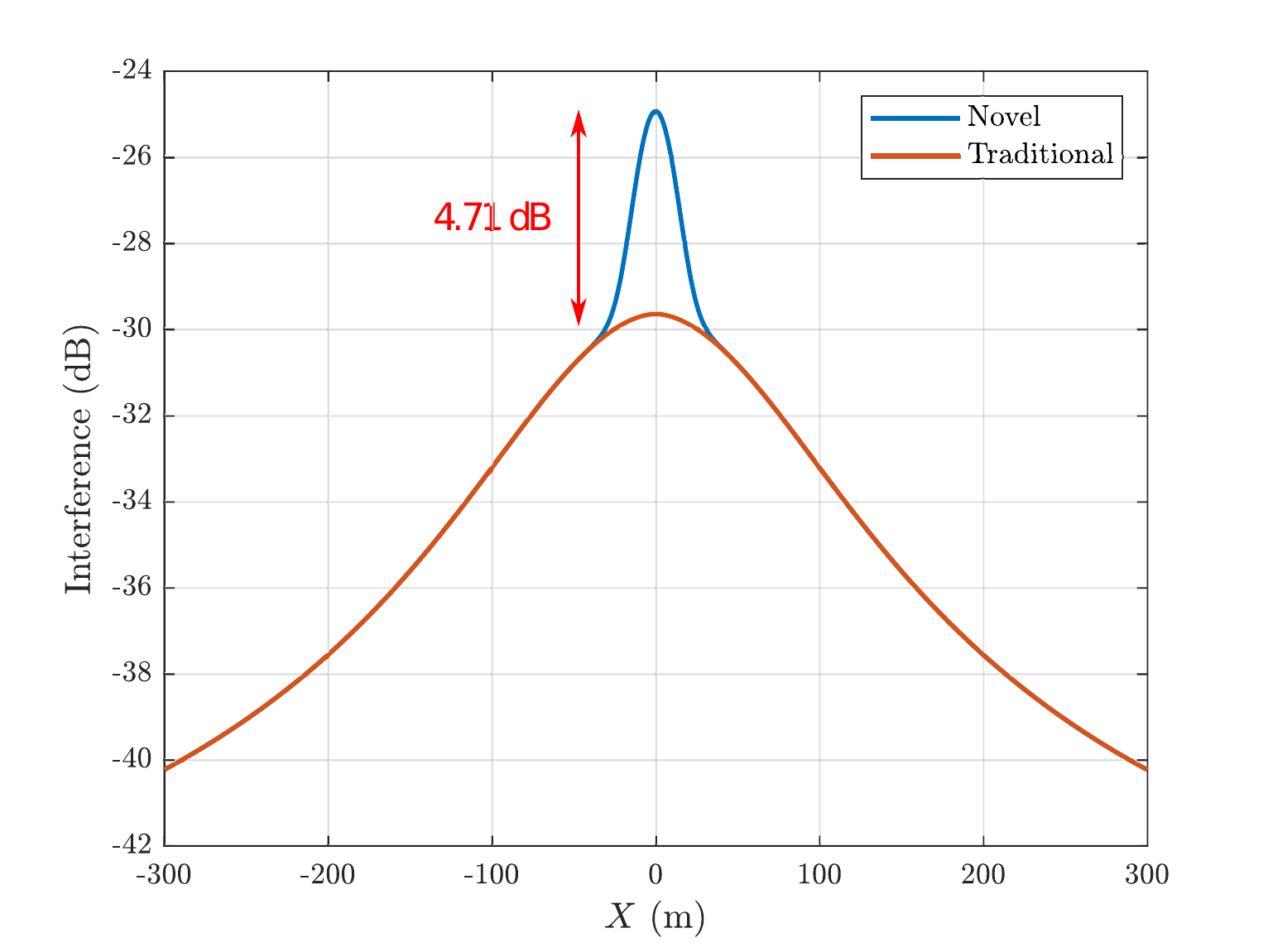}
        \caption{High-rise.}
        \label{fig:Interference_difference_highrise}
    \end{subfigure}
    \caption{Interference characterization of the traditional model and our model.}
    \label{fig:Interference_difference}
\end{figure*}

\section{EXTENDED MODEL EFFECT ON INTERFERENCE}
Equipped with this LOS model tailored for future UAM and smart vehicle environments, system designers can more effectively plan UAM infrastructure and air corridors to optimize coverage and minimize interference. Specifically, the extended model enhances the accuracy of predicting the received interference at NTNs and UAM aircraft from ground vehicles. We assume an NTN is flying at a height \(h=150~\text{m}\) perpendicular to a road, and the road has a vehicle density of one vehicle per \(10\)~m. We compute the interference for the extended (E) and the traditional (T) models as follows
\begin{equation}
    \begin{split}
        I^{(E,T)} & = \sum_{j\in \text{vehicles}} P_{\text{LOS}}^{(E,T)}\big(\theta_j\big) |g_{\text{LOS}}|^2 {(r_j^2+h^2)}^{-\frac{\alpha_{\text{LOS}}}{2}}\\
        & + \big(1-P_{\text{LOS}}^{(E,T)}\big(\theta_j\big)\big) |g_{\text{NLOS}}|^2 {(r_j^2+h^2)}^{-\frac{\alpha_{\text{NLOS}}}{2}},
    \end{split}
\end{equation}

where \(r_j\) is the ground distance between vehicle \(j\) and the NTN, and \(g_{\text{LOS}}\) and \(g_{\text{NLOS}}\) are small-scale fading channel  coefficients, following a Nakagami-\(m\) random variable with parameters \(m_{\text{LOS}}=1\) and \(m_{\text{NLOS}}=2\). Hence, the power coefficients will be Gamma distributed, i.e. \(|g_{\text{LOS}}|^2\sim\text{Gamma}(m_{\text{LOS}}, 1/m_{\text{LOS}})\) and \(|g_{\text{NLOS}}|^2\sim\text{Gamma}(m_{\text{NLOS}}, 1/m_{\text{NLOS}})\) \cite{cherif2020downlink}. Nakagami-\(m\) is used due to its generality in modeling various channel conditions \cite{aalo2015effect}. Moreover, \(\alpha_{\text{LOS}}\) and \(\alpha_{\text{NLOS}}\) are the path-loss exponents with values \(2\) and \(4\) respectively. Fig.~\ref{fig:Interference_difference} illustrates the discrepancy in interference power (in dB) received at an NTN when comparing the extended and traditional models. This shows that the traditional model significantly underestimates the received interference, particularly in densely populated, high-rise environments, where the disparity is most pronounced (up to \(4.71~\)dB). This critical insight underscores the importance of adopting the new model for more reliable UAM system design. Furthermore, by providing a more accurate interference characterization, the model can aid in minimizing unnecessary transmissions and optimizing resource allocation, thus reducing energy consumption. This is especially important for power-constrained aircraft equipment, where efficient interference management can improve connectivity and sustainability. The proposed model can also support next-generation communication networks, particularly fifth-generation (5G) UAM systems \cite{wanniarachchi2023study}. Given that 5G operates in millimeter-wave bands, which are highly sensitive to LOS conditions, integrating our model into 5G network designs can improve link reliability and beamforming accuracy.


\section{CONCLUSION}
In this paper, we proposed a modified road-aware LOS probability model designed to enhance the modeling accuracy of UAM communication systems and performance analysis. The model addressed limitations of the traditional model by enabling more accurate predictions of interference. We have rigorously tested our model against simulation data and validated it in diverse urban settings using real-world data. It is worth noting that the LOS model is based on a simulated uniform urban layout; however, for a more accurate representation, designers can incorporate localized building density functions of the environment in question. Additionally, while our model has been validated across difference urban topographies, its scalability in extremely-dense urban environments with highly complex layouts remains an important area for future research. For example, in cities with extreme vertical-density environments, factors such as sky bridges and multi-tier urban layouts can highly influence LOS conditions. Future work can also involve validating the model using real-world experimental measurements from UAM operations, which will improve the accuracy and applicability of the model. The model provides insights for understanding UAM network performance and for the strategic design of air corridors to minimize interference and optimize coverage. For example, our findings recommend configuring air corridors to run parallel to urban streets while maintaining a strategic distance to reduce cross-interference from ground vehicles and pedestrians. When designing air corridors as suggested, it is recommended to position NTN BSs along and above these corridors to significantly enhance coverage and reduce interference.

\begin{IEEEbiography}[{\includegraphics[width=1in,height=1.25in,clip,keepaspectratio]{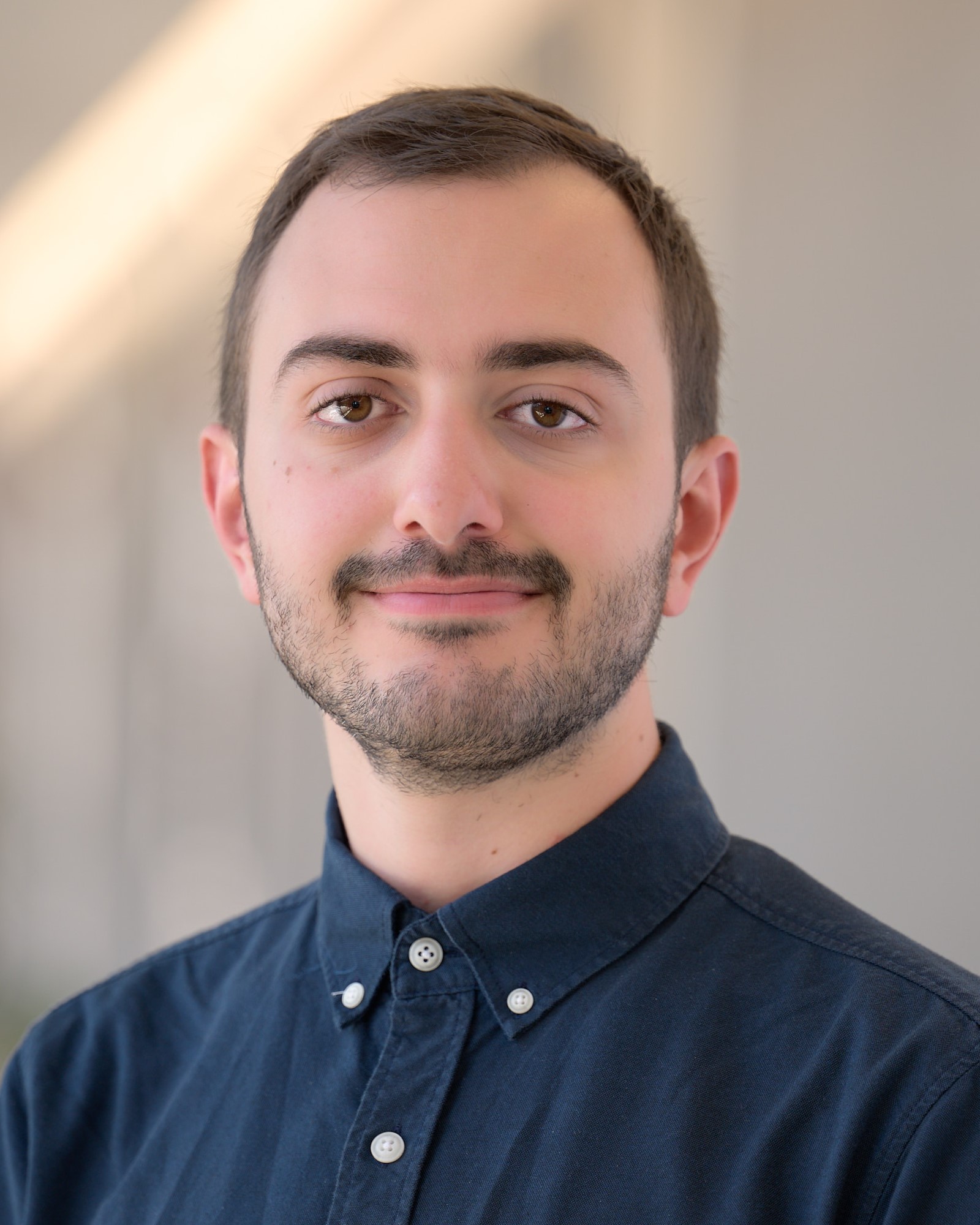}}]{Abdullah Abu Zaid}
(Graduate Student Member, IEEE) received the bachelor’s
degree from The University of Jordan, Amman,
Jordan, in 2020, and the master’s degree in electrical and computer engineering in 2021 from the
King Abdullah University of Science and Technology (KAUST), Saudi Arabia, where he is currently
working toward the Ph.D. degree with Communication Theory Laboratory (CTL), under the supervision of Professor Mohamed-Slim Alouini.
During his engineering studies, he did research in
wireless sensor networks and cognitive radio. His
research interests include stochastic geometry modeling, and flying platforms.
\end{IEEEbiography}

\begin{IEEEbiography}[{\includegraphics[width=1in,height=1.25in,clip,keepaspectratio]{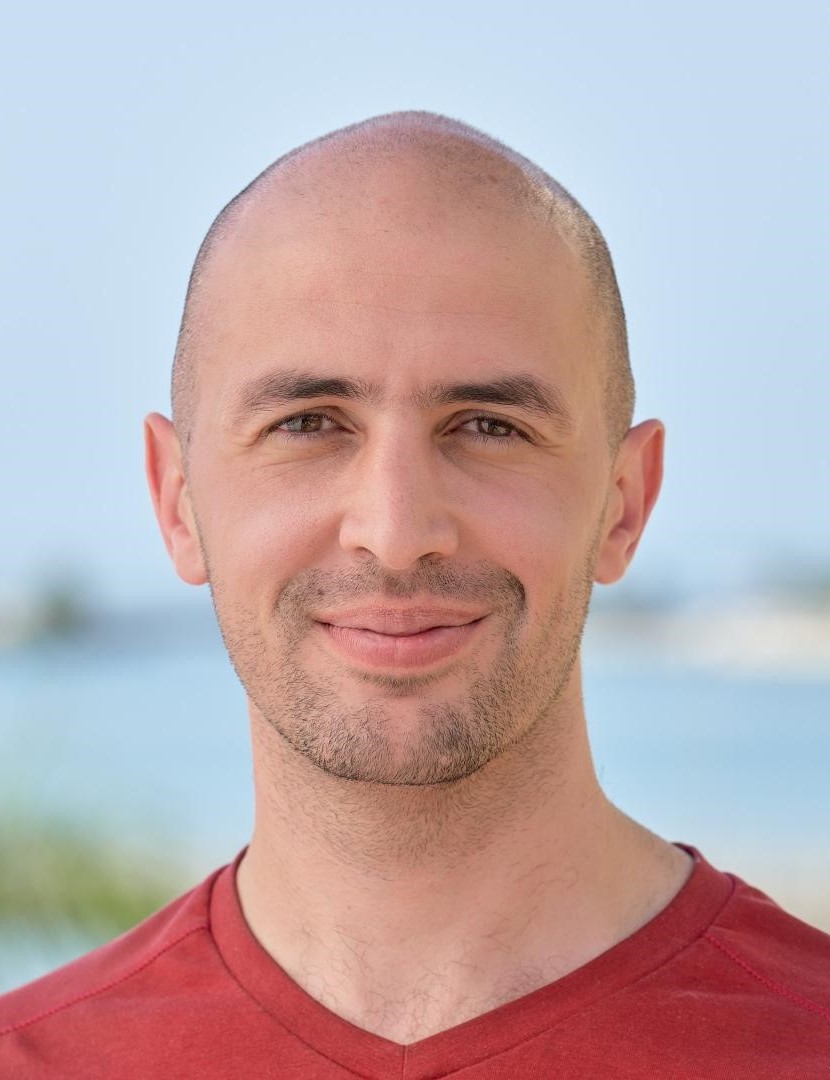}}]{Baha Eddine Youcef Belmekki} (Senior Member, IEEE) received his B.S. degree in Electrical Engineering and his M.Sc. degree in Wireless Communications and Networking from the University of Science and Technology Houari Boumediene (USTHB) in Algiers, Algeria, in 2011 and 2013, respectively. He completed his Ph.D. in Wireless Communications at the National Polytechnic Institute of Toulouse (INPT), Toulouse, France, in 2020. From 2013 to 2014, he worked as a Radio Access Network Engineer in Algiers, Algeria. Between 2014 and 2016, he served as a lecturer at USTHB. From 2019 to 2021, he was a teaching and research assistant at INPT. From 2021 to 2024, he was a postdoctoral research fellow at the Communication Theory Laboratory at King Abdullah University of Science and Technology (KAUST) in Thuwal, Saudi Arabia. He currently holds the position of Assistant Professor at Heriot-Watt University in Edinburgh, United Kingdom.

Throughout his career, he has received various awards and accolades. In 2024, he was honored with the Arab Scientific Community Organization award. He also received the Arab-American Frontiers Fellowship from the US National Academies of Sciences, Engineering, and Medicine. Additionally, he won the best paper award at the Global Advanced Air Mobility Academic Paper competition, organized by the International Civil Aviation Organization (ICAO). In 2023, he was the winner of the Falling Walls Lab and was recognized as an Emerging Talent at the Berlin Falling Walls. His research interests center around vehicular, advanced air mobility, and non-terrestrial networks.

\end{IEEEbiography}

\begin{IEEEbiography}[{\includegraphics[width=1in,height=1.25in,clip,keepaspectratio]{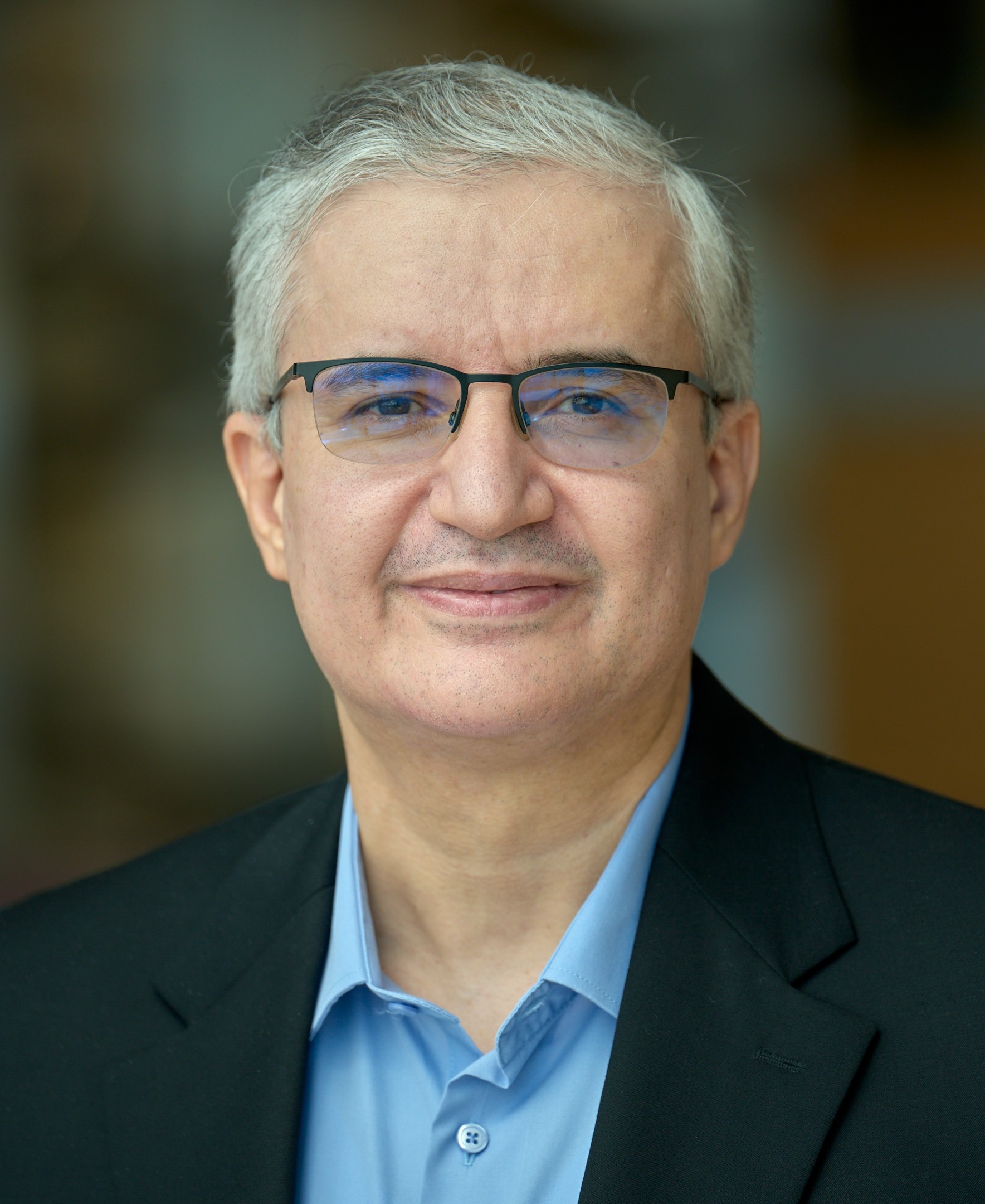}}]{Mohamed-Slim Alouini} (Fellow, IEEE) was born in Tunis, Tunisia. He earned his Ph.D. from the California Institute ofTechnology (Caltech) in 1998 before serving as a faculty member at the University of Minnesota and later at Texas A\&M University at Qatar. In 2009, he became a founding faculty member at King Abdullah University of Science and Technology (KAUST), where he currently is the Al-Khawarizmi Distinguished Professor of Electrical and Computer Engineering and the holder of the UNESCO Chair on Education to Connect the Unconnected. Dr. Alouini is a Fellow of the IEEE, OPTICA, and SPIE, and his research interests encompass a wide array of research topics in wireless and satellite communications. He is currently particularly focusing on addressing the technical challenges associated with information and communication technologies (ICT) in underserved regions and is committed to bridging the digital divide by tackling issues related to the uneven distribution, access to, and utilization of ICT in rural, low-income, disaster-prone, and hard-to-reach areas.
\end{IEEEbiography}

\end{document}